\documentclass[manuscript]{aastex}

\usepackage{natbib}
\usepackage{color}

\shorttitle{Solar wind turbulence at $0.72$ AU and solar minimum}
\shortauthors{Teodorescu et al.}

\begin{document}


\title{Inertial range turbulence of fast and slow solar wind at 0.72 AU and solar minimum}


\author{Eliza Teodorescu}
\affil{Institute for Space Sciences, M\u{a}gurele, Romania}
\email{eliteo@spacescience.ro}

\author{Marius Echim\altaffilmark{1}}
\affil{Belgian Institute for Space Aeronomy, Bruxelles, Belgium}

\author{Costel Munteanu\altaffilmark{2,3}}
\affil{Institute for Space Sciences, M\u{a}gurele, Romania}

\author{Tielong Zhang}
\affil{Space Research Institute, Graz, Austria}

\author{Roberto Bruno}
\affil{INAF-IAPS, Istituto di Astrofizica e Planetologia Spaziali, Rome, Italy}

\author{Peter Kovacs}
\affil{Geological and Geophysical Institute of Hungary, Budapest, Hungary}


\altaffiltext{1}{Institute for Space Sciences, M\u{a}gurele, Rom\^ania}
\altaffiltext{2}{University of Bucharest, M\u{a}gurele, Rom\^ania}
\altaffiltext{3}{University of Oulu, Oulu, Finland}
\altaffiltext{4}{Belgian Institute for Space Aeronomy, Bruxelles, Belgium}


\begin{abstract}
We investigate Venus Express (VEX) observations of magnetic 
field fluctuations performed systematically in the solar wind
at $0.72$ Astronomical Units (AU), 
between 2007 and 2009,  during the deep minimum of the solar cycle 24.
The Power Spectral Densities (PSD) of the magnetic field components have been computed
for the time intervals that satisfy data integrity criteria and have been grouped
according to the type of wind, fast and slow defined for speeds larger and respectively
smaller than 450 km/s. The PSDs show higher levels of power for the
fast than for the slow wind. The spectral slopes estimated for all PSDs in the frequency range
0.005-0.1 Hz exhibit a normal distribution. The average value 
of the trace of the spectral matrix
is $-1.60$ for fast  solar
wind and $-1.65$ for slow wind. Compared to the 
corresponding average slopes at 1 AU, the PSDs are 
shallower at 0.72 AU for slow wind conditions suggesting
a steepening of the solar wind spectra between Venus and Earth.
No significant time variation trend is observed for 
the spectral behavior of both slow and fast wind.

\end{abstract}


\keywords{interplanetary medium --- magnetic fields --- plasmas --- solar wind --- turbulence}



\section{Introduction}

The spectral properties of the  magnetic field and plasma fluctuations in the solar wind
have been investigated in-situ over several decades, for a broad range of frequencies and 
for various radial distances. It was found that the power spectral
density (PSD) of magnetic field fluctuations exhibits three different power law 
regimes, $P(k)=P_0 k^{-\alpha}$,  characterized by different exponents:  
(i)  $\alpha \approx -1$ for smaller $k$ (e.g. \citeauthor{MatthaeusGoldstein1986},
\citeyear{MatthaeusGoldstein1986}; \citeauthor{Verdini12},  \citeyear{Verdini12}), 
(ii) $\alpha \approx -5/3$ for the intermediate $k$ (e.g., \citeauthor{MarschTu1990}, \citeyear{MarschTu1990});
this range of $k$ is also anisotropic, the fluctuations parallel and perpendicular
to the magnetic field may exhibit different power law index (see, e.g., 
\citeauthor{Dasso2005}, \citeyear{Dasso2005}; \citeauthor{Horbury2012}, \citeyear{Horbury2012}) 
(iii) $ \alpha \le -2.5$ with a minimum index 
close to $-4.5$  \citep{Leamon1999,Bruno2014b}  for the largest $k$ 
 (see, also, \citeauthor{Coleman68}, \citeyear{Coleman68};  \citeauthor{Stawicki2001},
 \citeyear{Stawicki2001}, and \citeauthor{BrunoCarboneLRSP}, \citeyear{BrunoCarboneLRSP}, \citeauthor{Alexandrova2013}, \citeyear{Alexandrova2013}, for a review).
\citeauthor{Frisch95} (\citeyear{Frisch95}) described the three characteristic power law regimes, separated
by spectral breaks, as the magnetohydrodynamic equivalents of the scale ranges of
the classical hydrodynamic turbulence: (i) the driving (or energy containing)
range, (ii) the inertial range, dominated by  nonlinear turbulent
interactions that transfer the energy over multi-scales and (iii) the dissipation range. 
The physical processes contributing to dissipation in turbulent 
collisionless plasmas are still an open issue and in the recent years it has been argued (see, e.g., \citeauthor{Alexandrova2013}, \citeyear{Alexandrova2013}) that below the proton scales another turbulent cascade may take place that is described by a different power law. It is followed by an exponential law which could be indicative of dissipation.    

In practice, the analysis of in-situ time series provides PSD
as a function of the frequency in the spacecraft reference frame,  $P(f_{sat})$, that would correspond 
to Doppler shifted wavevector spectra, $P(k)$, under the assumption 
that the plasma flows over the spacecraft much faster than the 
characteristic time evolution  of the nonlinearly interacting turbulent spatial 
structures/eddies  (the Taylor hypothesis).
In the solar wind the transition between the driving  and 
the inertial range is in general observed at frequencies between 
$10^{-4}$ to  $10^{-3}$ Hz that would correspond to spatial scales related to
the  solar wind correlation/integral length ($\lambda$) 
or the typical size of the ``energy containing eddies''
\citep{Batchelor1970, Matthaeus1994}. The high frequency limit of the inertial range in the solar wind and the transition to kinetic regime is marked by a break in the spacecraft-frame frequency representation, generally in the vicinity of spatial scales (under the Taylor hypothesis) of the order of the proton inertial length or the proton Larmor radius (e.g.~\citeauthor{Chen2014},~\citeyear{Chen2014}). Recently, the variation of this break with the heliocentric distance was discussed  by~\citeauthor{Bruno2014a}, \citeyear{Bruno2014a}. High resolution data seem to suggest that dissipation may effectively start at higher frequencies, corresponding to the electron Larmor radius~\citep{Alexandrova2009}. The solar wind is a  supersonic and super-Alfv\'enic tenuous 
stream of collisionless plasma emerging from the dynamic solar corona 
therefore discerning ``pure'' turbulence features 
from other structures convected from the Sun
is  still an issue (see, for instance, \citeauthor{TuMarsch1995},
 \citeyear{TuMarsch1995}; \citeauthor{Brunoetal2007}, \citeyear{Brunoetal2007};
\citeauthor{Borovsky2008}, \citeyear{Borovsky2008}). 

Solar wind observations in the inner  heliosphere (between 0.3 and 0.9 AU) 
suggest that the ordering parameter
of turbulent properties is  the ``age'' of turbulence,
evaluated as the time it takes for the solar wind to travel from the
Sun to the spacecraft,  rather than the radial distance.
The ``aging'' of the solar wind turbulence is also characterized
by a progressive spectral dominance of the
``2D'' mode of  turbulence (characterized mainly by perpendicular wavevectors),
over the ``slab'' mode (dominated mainly by
parallel wavevectors)  \citep{Ruiz2011}.
Analyses based on a global mean magnetic field estimation concluded that the slow wind exhibits in general features of ``2D''  turbulence while the turbulence in the fast wind is more
of the ``slab'' type \citep{Dasso2005, Weygand2011} and the anisotropic state found near the Sun evolves towards a more isotropic state at 1 AU. On the other hand, approaches exploring anisotropy through the scale-dependent local mean magnetic field (e.g.~\citeauthor{Horbury2008}, \citeyear{Horbury2008}, \citeauthor{Podesta2009}, \citeyear{Podesta2009}, \citeauthor{Wicks2010}, \citeyear{Wicks2010}, \citeauthor{Forman2011}, \citeyear{Forman2011}) indicate that the high-speed solar wind power spectrum is dominated by perpendicular ``2D'' fluctuations. Simulation results~\citep{Chen2011} show the same discrepancy between global and local mean magnetic field approaches and the authors conclude that the global mean magnetic field scaling is not able to properly discriminate between parallel and perpendicular fluctuations. \citeauthor{Smith2003} (\citeyear{Smith2003}) shows that at high latitudes roughly equal proportions of “slab” (1D) and “2D” coexist in the same plasma element.


At solar minimum the solar wind is characterized by an increased recurrence of 
the high speed streams (up to 800 km/s and more),
lower density and higher temperature, whose origin is the meridional extensions of the polar coronal holes.   
The properties of fast and slow wind turbulence were investigated
for different phases of the solar cycle, 
from data recorded by Helios \citep{Bavassano1989, Bavassano1991, Ruiz2011}, 
ACE \citep{Borovsky2012b}, Ulysses  \citep{Yordanova2009}, 
Cluster and THEMIS \citep{Weygand2011}, Messenger and Wind \citep{Bruno2014a, Bruno2014b}. 
Observations of the solar wind by Ulysses at larger radial distances (between $1.5$ AU and $5.4$ AU),
outside the ecliptic and close to the solar minimum show that 
the spectral index of the fast wind inertial range turbulence
takes values in the range $-1.79<\alpha<-1.55$ for the magnetic field components,
and between $-1.52<\alpha<-1.25$ for the total field, $|B|$. 
The spectra of the slow wind exhibit similar power law behavior but with steeper slopes, 
$-1.95<\alpha<-1.45$ for the components of the magnetic 
field and $-1.78<\alpha<-1.55$ for the total field \citep{Yordanova2009}.
It is not clear whether the inertial range spans roughly 
the same frequency range for the fast and slow wind.

In-situ observations at 1 AU indicate that
the \textit{median} of the magnetic spectral index in the inertial range
depends on the type of wind: it is shallower for the fast wind ($V_{wind}>450$ km/s), $\overline{\alpha}=-1.54$, compared
to the slow wind, $\overline{\alpha}=-1.70$, as shown
by \citet{Borovsky2012a} from ten years of ACE data (1998-2008). 
The spectral index may take ``extreme'' values, larger 
than $-1.33$ and smaller than $-1.95$. 
The steeper spectral slopes are observed at 1 AU
when the solar wind density is larger, the temperature is smaller,
the speed takes smaller values and the number of strong directional 
discontinuities is reduced  \citep{Borovsky2012a}. 
Close to the high frequency limit of the inertial range, at proton scales,
in the vicinity of the fluid/kinetic spectral break, 
WIND (at $0.99$ AU) and Messenger (at $0.42$ AU) data show that the spectral slope 
may depend on the power density in the inertial range:
steeper slopes are observed for larger power in the inertial subrange
\citep{Bruno2014b}. This frequency break moves towards smaller frequencies as the radial distance increases \citep{Bruno2014a}.

\section{Spectral properties of fast and slow solar wind at $0.72$ AU}

We analyze data recorded by Venus Express (VEX) in the solar wind,
in the vicinity of Venus, at 0.72 AU, between January 2007 and December 2009,
during the minimum of the solar cycle 24. Since Venus has no intrinsic
magnetic field its induced magnetosphere is confined to shorter distances from the
planet, thus VEX spends in the solar wind more than 20 hours each
day of the year. Thus VEX is a unique solar wind monitor that investigates 
the inner heliospheric solar wind on a day to day basis for almost one solar cycle,
since 2006.
The turbulent fluctuations of the solar wind magnetic field 
considered in this study are provided by Venus Express Magnetometer, 
VEX-MAG, \citep{Zha06}) with a cadence of 1 Hz. The data are obtained through an offline calibration procedure  
by donwsampling the 32 Hz resolution data. 

The plasma state (electron and ion spectra and their moments, e.g. 
density, temperature, velocity) is investigated
by  the Analyzer of Space Plasma and Energetic Atoms (ASPERA,  
\citeauthor{Bar07}, \citeyear{Bar07}).
ASPERA operates in the solar wind for short time intervals of the order
of one to two hours, close to the orbit apogee. The ion and electron
spectra and their moments are provided with a time resolution of 196 seconds
which does not allow a spectral analysis of the solar wind plasma parameters.
Nevertheless, the estimation of the moments of the ion velocity 
distribution function provides 
the data needed to select high and low speed solar wind. 
Inspired by previous studies, we select the slow and fast wind intervals based
on a threshold speed value equal to $450$ km/s.
The magnetic field experiment is operating continuously, 
however we consider time intervals of roughly four hours length, close to the
VEX apogee and that include the time periods when ASPERA is also operating.

\begin{figure}[!htbp]
\plotone{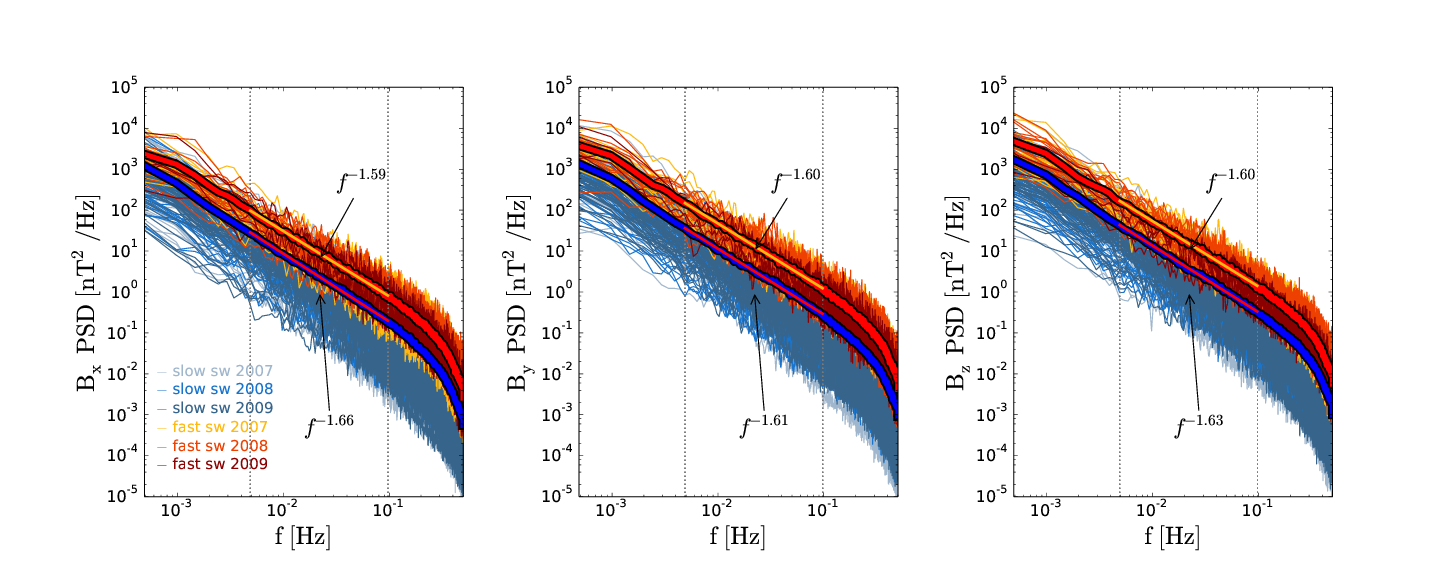}
\caption{Summary plot of the power spectral density spectrum of $\mathrm{B_x}$, $\mathrm{B_y}$ and $\mathrm{B_z}$ 
from VEX magnetic field data recorded between 2007-2009, during the
minimum of the 24th solar cycle. Yellow and red lines correspond to fast solar wind, gray and blue lines correspond to slow solar wind. 
The black dotted lines mark the frequency interval for which 
 the spectral index of each spectrum is computed. 
An average spectrum is derived as an ensemble average of all the spectra obtained for
fast wind (the embossed red line) and respectively the slow wind (the embossed blue line). The spectral slope of the
average spectra is computed for the frequency range marked by yellow and magenta respectively.}
\label{fig:sumplotBxyz}
\end{figure}

Additional constraints on the magnetic field data analysis
are imposed by several quality checks. We disregard the time intervals shorter than one hour
and that have data gaps whose largest size exceeds 30 consecutive points.  
In general, the total missing data points amount to less
than 3\% of the total number of samples for the  selected time intervals and they appear to be randomly distributed over time. 
Thus, the number of data gaps in the selected time intervals is small 
and their length is short. A linear interpolation is applied prior to the spectral analysis.
From a total of 1094 orbits between January 2007 and December 2009, only 204 time intervals
fulfill the data quality requirements, of which 48 time intervals correspond to fast 
solar wind observations ($V_{wind}>450$ km/s).

The magnetic field components are provided in the 
Venus Solar Orbital (VSO) rectangular frame, with the Ox axis aligned in the Sunward direction and
the Oz axis perpendicular to the ecliptic plane, in the Northward direction.
The PSDs are computed  with a Welch  algorithm~\citep{Wel67} of averaging periodograms
applied on $\mathrm{B_x}$, $\mathrm{B_y}$, 
$\mathrm{B_z}$, and the total field, $|B|$ for all the selected time intervals.
We also compute the trace of the spectral matrix of the fluctuations.
A summary plot of all the power spectra is given 
in Fig.~\ref{fig:sumplotBxyz};
different colors illustrate different types of wind (fast/slow) and different years. 
An average spectrum is computed as an ensemble average of all the spectra,
for the fast and the slow solar wind, respectively. 

\begin{figure}[!htpb]
\plotone{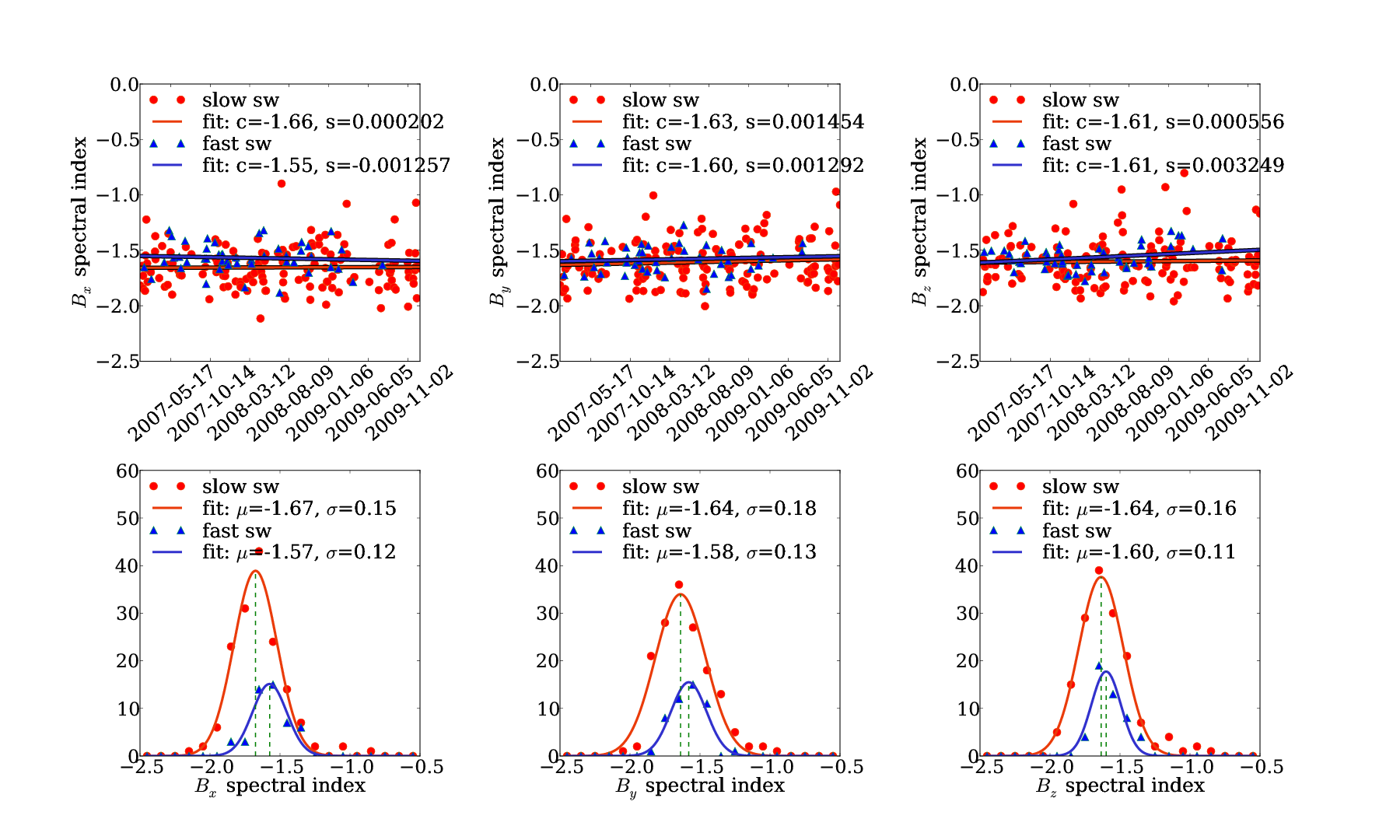}
\caption{The distribution of spectral index for $\mathrm{B_x}$, $\mathrm{B_y}$ and $\mathrm{B_z}$
from data recorded between 2007-2009. 
Upper raw: evolution in time of the spectral indices computed in the frequency interval indicated in Fig.~\ref{fig:sumplotBxyz}. 
{\it c} is the constant (intercept) of the first degree polynomial linear fit and {\it s} is the slope of the fit. 
Lower raw: the histogram of the spectral indices approaches a Gaussian distribution with a mean close to  $-5/3$ for
 the slow wind and $-1.6$ for the fast wind.}
\label{fig:spectral}
\end{figure}

The spectral power of fast solar wind is systematically larger than for
slow wind as shown by previous results in other locations of the heliosphere (see the review by~\citeauthor{BrunoCarboneLRSP},~\citeyear{BrunoCarboneLRSP} and recent results at 0.38 AU and 1 AU  by~\citeauthor{Bruno2014b},~\citeyear{Bruno2014b}). The power spectral densities 
exhibit a power law regime in the frequency range $[5\times 10^{-3}, 10^{-1}]$ Hz. 
A change in the spectral slope is observed around 0.2-0.3 Hz, close to the local Doppler shifted proton gyrofrequency, 
followed by a frequency range showing a steepening of the spectra that could 
correspond to the lowest frequencies of the non-MHD kinetic turbulent cascade. 
In order to minimize the effect of inherent uneven sampling of the frequency range
by the periodogram technique, the spectral index/slope has been computed from
PSDs rebinned and averaged over equal logarithmic bins of frequency. 
In addition to the rebinning of the frequency range we have also
searched for the existence of the power law behavior 
by varying the limits of the fitting frequency range. Thus for each of the power spectra the spectral index (or slope) is calculated 
by a linear least-square fitting over the  interval  $[5\times 10^{-3}, 10^{-1}]$ Hz.
This range would correspond to the high-frequency part of the inertial subrange at 0.72 AU and 
solar minimum.
The amount of data intervals is sufficiently large to allow a statistical analysis of 
the distribution of the spectral indices obtained at solar
minimum, between 2007 - 2009, as illustrated by Fig.~\ref{fig:spectral}.

\begin{figure}[!htpb]
\plotone{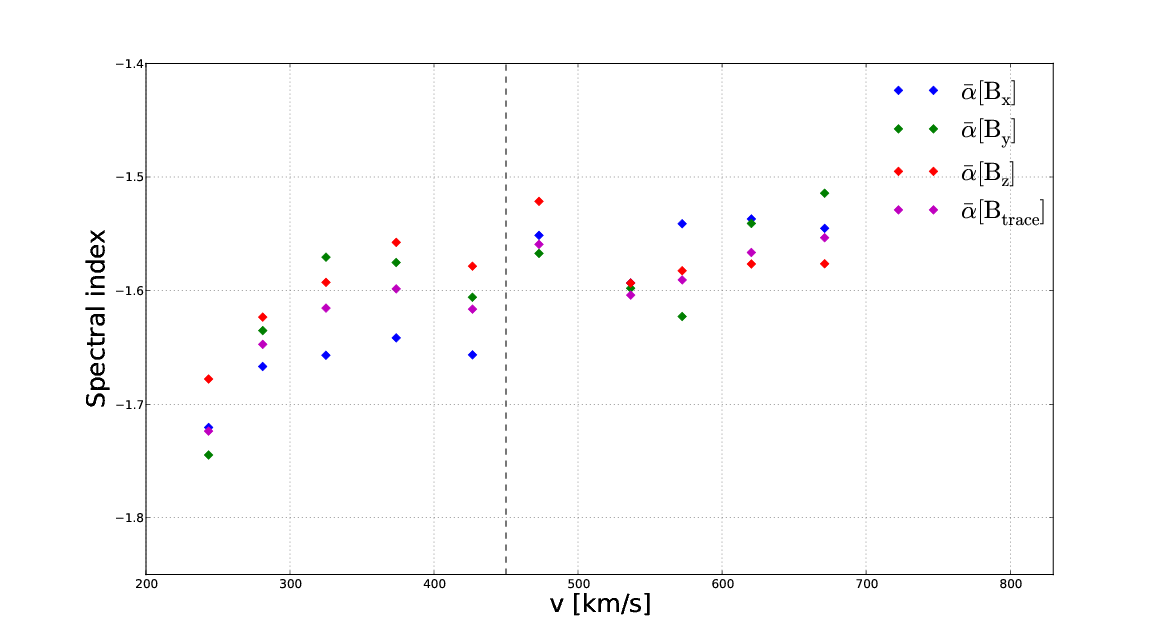}
\caption{The spectral index as a function of solar wind speed; we show averages over velocity bins of 50 km/s width. Different colors ilustrate different magnetic field components. Data is collected in the solar wind by VEX between 2007-2009. The dotted vertical line represents the threshold we chose to select fast and slow solar wind.} 
\label{fig:velindex}
\end{figure}

The {\it a priori} separation of fast from slow solar wind allows 
the independent and simultaneous tracking of the evolution of 
the spectral properties of turbulence for the two types of wind. 
The least-squares linear fitting of the distributions of the spectral slopes
(upper panels of Fig. \ref{fig:spectral})
yields an experimental evidence for the fact that the spectral indices do not exhibit any temporal trends 
over the years while the solar minimum deepens, both for slow and fast solar wind.
This can be considered an indication that the processes that contribute to the power law scaling
are mainly local, possibly related to non-linear interactions leading to
turbulent transfer of the energy between scales.
The histograms of the spectral indices show that the slopes of the magnetic field power spectra 
in the inertial range fit a Gaussian distribution, 
as seen in the lower row of plots in Fig.~\ref{fig:spectral}.
The  three components of the magnetic  field exhibit different average 
spectral slopes for fast and slow wind: $\overline{\alpha}_x^{fast}=-1.57\pm0.02$,
 $\overline{\alpha}_y^{fast}=-1.58\pm0.02$,  $\overline{\alpha}_z^{fast}=-1.60\pm0.02$ and
respectively  $\overline{\alpha}_x^{slow}=-1.67\pm0.01$,
 $\overline{\alpha}_y^{slow}=-1.64\pm0.01$,  $\overline{\alpha}_z^{slow}=-1.64\pm0.01$, where $\overline{\alpha}_i^{fast/slow}$ is 
the notation for the mean spectral index of the component $i$ and fast or slow type of wind.
The mean spectral index of the trace of the spectral matrix (not shown) varies
from  $\overline{\alpha}_{B}^{fast}=-1.60\pm0.01$ in the fast wind to 
 $\overline{\alpha}_{B}^{slow}=-1.65\pm0.01$ in the slow wind.

The variation of the spectral index as a function of the solar wind speed is shown in Fig.~\ref{fig:velindex}. The average spectral indices of the magnetic field become shallower with increasing plasma velocity which is in good agreement with previous work~\citep{Che13}. The continuous trend also seems to indicate that the solar wind velocity may be a controlling parameter for the spectral slope. Nevertheless, recent studies indicate that cross-helicity is equally important in controlling the spectral behavior~\citep{Che13}.

The Venus Express results show similarities with the spectral analysis of WIND 
magnetic field data at 1 AU, during roughly the same time interval,
between June 2004 and April 2009. Data from WIND showed a mean spectral slope 
between $-1.6$ (for wind speeds larger than $600$ km/s) to
$-1.72$ (for speeds smaller than $400$ km/s) \citep{Che13}. 
Another analysis of solar wind turbulence at 1 AU, based on ACE data recorded
between 1998-2008, suggests a median spectral slope of
the magnetic field equal to $-1.54$ for speeds larger than
$550$ km/s and respectively $-1.70$ for speeds smaller than $450$ km/s
\citep{Borovsky2012a}. Nevertheless, the Venus Express data suggest that
in average the slope of the slow wind spectra
are shallower at 0.72 AU than at 1 AU, 
while the fast wind shows in average comparable slopes.
We note the differences between the mean spectral slopes of the three
magnetic field components for slow wind conditions and roughly
the same mean slope for fast wind. On the other hand more
spectral power is observed for $B_y$ and $B_z$ components in fast wind.
This is in agreement with expectations of finding more power on the perpendicular components than the parallel one as fast wind is more Alfv\'{e}nic and Alfv\'{e}nic fluctuations are mainly perpendicular to the local mean magnetic field and are not compressive.
At 0.72 AU the Parker spiral (mean field) for fast (700km/s) solar wind would be around 24$^{\circ}$, not far from the X direction in the VSO reference system. Thus, more power should be expected on the Y and Z components of the magnetic field~\citep{Klein1993}.
The anisotropy will be the subject of a future study on the same dataset.

\section{Discussion and Conclusions}

We have investigated the spectral behavior of the solar wind 
magnetic field at 0.72 AU, for low solar activity, 
between 2007 and 2009, using data provided by Venus Express. 
The Power Spectral Densities of magnetic 
field components and the trace of the spectral matrix indicate that 
an inertial range of turbulence can be identified as a power-law behavior,
in the fast and slow solar wind. More power is contained in the
fast wind spectra that also exhibit shallower slopes 
than the slow wind.   
The mean value of the slope takes values around 
$-1.6\pm0.01$ for the fast wind and respectively $-1.65\pm0.01$ for the slow wind. 
Our results fully agree with general predictions found in literature about different spectral slopes of magnetic field fluctuations depending on solar wind conditions.

In  particular,~\citep{Che13} clearly showed how the spectral index of magnetic field and velocity fluctuations depend on their Alfv\'{e}nicity which can be expressed by the normalized cross-helicity $\sigma_C$. These authors, analysing 5 years of WIND data, found that as $\sigma_C$ decreases, magnetic energy starts to dominate on kinetic energy and the magnetic field spectrum becomes steeper than velocity spectrum, in line with predictions~\citep{MullerGrappin2004}, among others. The tendency for the magnetic spectrum to dominate over the kinetic one is a natural outcome for a stationary critically balanced MHD turbulence~\citep{Gol95} generated by non-linear interacting Alfv\'{e}n waves~\citep{Gogoberidze2012}.
As a consequence, since we study fast and slow wind, which differ in Alfv\'{e}nicity~\citep{BrunoCarboneLRSP}, fast wind being more Alfv\'{e}nic than slow wind, we should expect to find magnetic field spectra in fast wind less steep than those in slow wind. In addition, a comparison with similar data at 1 AU suggests that in average the spectra
steepen while the slow solar wind is transported between 0.72 and 1 AU suggesting that non-linear interactions are at work. 
On the other hand, the fast wind exhibits less clear evidence of the radial steepening.

The solar wind magnetic field spectral indices between 2007 and 2009 
have a normal distribution. We do not find a significant  temporal trend
of the slopes. The average spectral slopes
of the three magnetic components suggest an anisotropic repartition
of power. 
A change in the spectral slope is evidenced 
in the vicinity of the proton cyclotron radius, possibly 
associated to the transition from the inertial to the
kinetic subrange.

Our results suggest that at 0.72 AU and solar minimum the slow wind exhibits in average a spectral behavior closer to the ``$f^{-5/3}$'' law with some differences between the mean slopes of the magnetic field components. Thus there are indications that the turbulence is anisotropic and models based on the isotropy hypothesis, like Kolmogorov and/or Iroshnickov-Kraichnan, are not applicable. Modern theories of anisotropic strong MHD turbulence~\citep{Gol95} predict that the perpendicular spectrum of turbulence may approach in some conditions the ``$f^{-5/3}$'' scaling. In the absence of resolute plasma measurements from VEX we can only suggest that the slow wind spectra are perhaps dominated by the perpendicular component. Nevertheless the ”younger” turbulence carried by the fast wind is described by shallower spectral slopes that show a tendency to approach asymptotically a ``$f^{-3/2}$'' power law. This could possibly signify that the structure of turbulence exhibits features consistent with models of anisotropic turbulence like in~\citeauthor{Bol2006} (~\citeyear{Bol2006}). In other words, our results may suggest that the slow wind at 0.72 AU and solar minimum is dominated by filament-like structures at the smallest scales (as suggested by~\citeauthor{Gol95},~\citeyear{Gol95}), while the fast wind turbulence is dominated by the sheet-like structures, at the smallest scales (as suggested by~\citeauthor{Bol2006},~\citeyear{Bol2006}), possibly related to phenomenological turbulent features of the solar wind at the origin, in the corona.




\acknowledgments

Research supported by the European Community’s Seventh Framework Programme (FP7/2007-2013) under grant agreement no 313038/STORM, 
and a grant of the Romanian Ministry of National Education, CNCS – UEFISCDI, project number PN-II-ID-PCE-2012-4-0418. 
Partially, data analysis was done with the AMDA science analysis system provided by the Centre de Données de la Physique 
des Plasmas (IRAP, Université Paul Sabatier, Toulouse) supported by CNRS and CNES.


{\it Facilities:} \facility{Venus Express}.

\clearpage



\end{document}